\documentclass[useAMS,usenatbib]{mn2e}

\usepackage{epsfig}
\usepackage{graphics}
\usepackage{afterpage}
\usepackage{times}
\usepackage{amssymb}
\usepackage{amsmath}

\newlength{\colwidth}
\newcommand{\ion}[2]{\hbox{#1\,{\sc #2}}}

\newcommand{\aap}{A\&A}
\newcommand{\apj}{ApJ}

\newcommand{\mnras}{MNRAS}

\newcommand{\apjs}{ApJS}

\title[Calibrating Galaxy Redshifts Using the IGM]{Calibrating Galaxy Redshifts Using Absorption by the Surrounding Intergalactic Medium\thanks{Based on data obtained at the W.M. Keck Observatory, which is operated as a scientific partnership among the California Institute of Technology, the University of California, and NASA, and was made possible by the generous financial support of the W.M. Keck Foundation.}} 

\author[O. Rakic et al.]{%
Olivera~Rakic,$^1$\thanks{E-mail: rakic@strw.leidenuniv.nl} 
Joop~Schaye,$^1$ Charles~C.~Steidel,$^2$ 
and Gwen~C.~Rudie$^2$\\  
\\
$^{1}$ Leiden Observatory, Leiden University, P.O. Box 9513, 2300 RA Leiden,
  the Netherlands\\
$^2$ California Institute of Technology, MS 249-17, Pasadena, CA 91125, USA\\
}

\begin{document}

\pagerange{\pageref{firstpage}--\pageref{lastpage}} \pubyear{2007}

\maketitle

\label{firstpage}

\begin{abstract}
Rest-frame UV spectral lines of star-forming galaxies are systematically offset from the galaxies' systemic redshifts, probably because of large-scale outflows. We calibrate galaxy redshifts measured from rest-frame UV lines by utilizing the fact that the mean \ion{H}{I} Ly$\alpha$ absorption profiles around the galaxies, as seen in spectra of background objects, must be symmetric with respect to the true galaxy redshifts if the galaxies are oriented randomly with respect to the lines of sight to the background objects. We use 15 bright QSOs at $z\approx2.5-3$ and more than 600 foreground galaxies with spectroscopic redshifts at $z\approx1.9-2.5$. All galaxies are within 2 Mpc proper from the lines of sight to the background QSOs. We find that  Ly$\alpha$ emission and ISM absorption redshifts require systematic shifts of $\Delta v_{\rm Ly\alpha}=-295_{-35}^{+35}\rm \, km\, s^{-1}$ and $\Delta v_{\rm ISM}=145_{-35}^{+70}\rm\, km\, s^{-1}$, respectively. Assuming a Gaussian distribution, we put 1$\sigma$ upper limits on possible random redshift offsets of $<220\rm \, km\, s^{-1}$ for Ly$\alpha$ and $< 420 \rm \, km\, s^{-1}$ for ISM redshifts. For the small subset ($< 10$ percent) of galaxies for which near-IR spectra have been obtained, we can compare our results to direct measurements based on rest-frame optical, nebular emission lines, which we confirm to mark the systemic redshifts. While our $\Delta v_{\rm ISM}$ agrees with the direct measurements, our $\Delta v_{\rm Ly\alpha}$ is significantly smaller. However, when we apply our method to the near-IR subsample which is characterized by slightly different selection effects, the best-fit velocity offset comes into agreement with the direct measurement. This confirms the validity of our approach, and implies that no single number appropriately describes the whole population of galaxies, in line with the observation that the line offset depends on galaxy spectral morphology. This method provides accurate redshift calibrations and will enable studies of circumgalactic matter around galaxies for which rest-frame optical observations are not available. 
\end{abstract}

\begin{keywords}
galaxies: high-redshift  --- intergalactic medium --- quasars: absorption lines
\end{keywords}

\section{Introduction}

Two questions that are key to understanding galaxy formation and evolution are: How do galaxies get their gas? and Where does the gas that is ejected via galactic feedback processes end up? A powerful method to study the interface between galaxies and the intergalactic medium (IGM) is to examine the spectra of bright background objects for absorption features at the redshifts of foreground galaxies. Different studies have looked into the relation between Ly$\alpha$ absorbing gas and galaxies \citep[e.g.,][]{Lanzetta1995,Chen1998,Bowen2002,Penton2002,Adelberger2003,Adelberger2005,Crighton2010,Steidel2010}, as well as metals and galaxies \citep[e.g.,][]{Lanzetta1990,Bergeron1991,Steidel1994,Steidel1997,Chen2001,Adelberger2003,Adelberger2005,Pieri2006,Steidel2010}.  A crucial requirement for the  success of such studies is the availability of accurate redshifts for the foreground galaxies. To illustrate this point, we note that observationally inferred outflow velocities correspond to a change in redshift of order  $\Delta z/(1+z)=\Delta v/c\sim 10^{-3}$ and that at $z=2$ the difference
in Hubble velocity across a proper distance of 1~Mpc, which exceeds the virial radius
of the typical star-forming galaxy at that redshift by more than an
order of magnitude, corresponds to $\Delta z/(1+z) \approx 7\times 10^{-4}$. Clearly, the systematic errors on the redshifts of foreground objects need to be $\ll0.1\%$ in order to map the physical properties of the gas in and around the haloes of galaxies.

The most accurate galaxy redshifts are measured from  absorption lines arising in stellar atmospheres. Rest-frame optical stellar absorption features are routinely detected in spectra of galaxies in the local Universe. For high-$z$ galaxies rest-frame UV stellar absorption lines are detectable in stacks of $\approx100$ high-quality spectra and have been used to verify the accuracy of redshift calibrations \citep{Shapley2003,Steidel2010}. One can also measure accurate redshifts from nebular emission lines from stellar HII regions. Lines such as H$\alpha$ ($\lambda$6563), H$\beta$ ($\lambda$4861), and  [\ion{O}{III}] ($\lambda\lambda$4959, 5007) are strong and easily accessible for galaxies in the nearby Universe because they are in the rest-frame optical. Measuring redshifts from these lines  is, however, more difficult for high-$z$ galaxies. Nebular lines are redshifted into the observed frame near-infrared, and at these wavelengths spectroscopy with ground based instruments is complicated by the bright night-sky, as well as by strong absorption features produced by molecules in the Earth's atmosphere. The prospects for near-IR spectroscopy, at least for bright objects, will improve when multi-object near-IR spectrographs come online \citep[e.g.\, Keck I/MOSFIRE; ][]{McLean2008}, but near-IR spectroscopy will remain challenging for fainter objects.

Alternatively, one may resort to measuring redshifts using spectral lines that lie in the rest-frame UV and can thus be observed with optical spectrographs. Such  lines include the \ion{H}{I} Ly$\alpha$ emission line at 1216\AA, when present, and UV absorption lines arising in the interstellar medium, of which the strongest are \ion{C}{II} ($\lambda$1335), \ion{Si}{IV} ($\lambda$1394), and \ion{Si}{II} ($\lambda$1527). However, these lines are not ideal as they do not provide accurate redshift measurements. Ly$\alpha$ photons are resonantly scattered, and the observed line profile depends on the kinematics and the amount of gas that is scattering the photons, as well as on the dust content \citep[e.g.,][]{Verhamme2006,Verhamme2008,Shaerer2008,Hansen2006,Dijkstra2006a,Dijkstra2006b,Zheng2002}. The observed velocities of interstellar absorption lines also depend on the kinematics of the gas that produces them. Most high-redshift galaxies observable with current facilities have sufficiently high star formation rates to drive galactic scale winds. These winds carry some interstellar material out of the galaxies and this causes both the Ly$\alpha$ emission line and interstellar absorption lines to be systematically offset by a few hundred $\rm km\, s^{-1}$ from the galaxy systemic redshifts, as measured from nebular lines \citep[e.g.,][]{Pettini2001, Adelberger2003, Adelberger2005, Steidel2010}.

In surveys of high-redshift ($z\approx2-3$) star-forming galaxies, \citet{Adelberger2003,Adelberger2005} and \citet{Steidel2010} dealt with these issues by measuring redshifts from rest-frame UV lines, obtaining near-IR spectroscopy for a subset ($<10\, \%$) of galaxies, and then calibrating the redshifts measured from rest-frame UV lines using the rest-frame optical nebular lines. This calibration was then applied to the rest of the sample that lacked near-IR spectroscopy. However,  galaxy populations other than Lyman Break Galaxies (LBGs) may require their own calibrations. This is a problem as calibration using nebular lines is often not possible. For example, most of the galaxies found by future surveys with integral field spectrographs, such as the Multi Unit Spectroscopic Explorer \citep[MUSE,][]{Bacon2010}, will only be detected in Ly$\alpha$. Also, even for the LBGs at $z\approx2-3$ it is possible that the subsample that is observed in the near-IR is not quite representative of the whole sample, for example because galaxies need to be sufficiently bright to do near-IR spectroscopy.

In this paper we present and apply a method for calibrating redshifts measured from rest-frame UV lines using the absorption features from the surrounding IGM as seen in the spectra of background QSOs. Several studies have found that the mean strength of \ion{H}{I} Ly$\alpha$ absorption increases as the distance to the nearest galaxy decreases \citep[e.g.,][]{Chen1998,Adelberger2003, Adelberger2005,Pieri2006,Ryan-Weber2006,Morris2006,Wilman2007,Steidel2010}. By using the fact  that the mean absorption profiles must be symmetric with respect to the systemic galaxy redshifts if the galaxies are oriented randomly with respect to the lines of sight to the background QSOs, we determine  the optimal average velocity offset for which the observed mean absorption profile is symmetric around the positions of the galaxies  along the line of sight (LOS). Although the same principle was used to verify the accuracy of the calibrations proposed in \citet{Adelberger2003} and \citet{Steidel2010}, this is the first time that the IGM absorption is used for calibration itself. The use of absorption by the IGM makes the method independent of the galaxies' luminosities, and does not require rest-frame optical observations. We find that the required offsets for redshifts based on ISM absorption lines and Ly$\alpha$ emission lines agree with the calibrations inferred by direct comparison with nebular line redshifts. Using this method we also confirm that different galaxy subsamples require different calibrations depending on the galaxies' spectral morphology.  

Throughout this work we use $\Omega_{\rm m}=0.258$, $\Omega_{\Lambda}=0.742$, and $h=0.719$ \citep{Komatsu2009}. Unless stated otherwise, all distances are proper. 

\section{Data}
\label{Data}

The data sample used in this study is partially (3 out of 15 fields) described in Steidel et al. (2003; 2004), \citet{Adelberger2005}, and \citet{Steidel2010}, while the full survey will be presented elsewhere. Here we give only a brief description. This redshift survey was constructed to select galaxies whose $U_{n}GR$ colors are consistent with star-forming galaxies at redshifts $z\approx 1.9-2.7$. The survey was conducted in 15 fields with typical size $5\times7\, \rm arcmin$, centred on bright background QSOs for which there were high-resolution spectra suitable for probing Lyman-$\alpha$ absorbing gas in the same redshift range. The technique for optically selecting galaxies is described in detail in Adelberger et al. (2004) (their BX/BM sample). All 659 galaxies used here have been followed up spectroscopically with the optical spectrograph Keck I/LRIS-B \citep[FWHM $\approx370\,\rm km\, s^{-1}$; ][]{Steidel2004}. A subset of 49 galaxies  has been observed with the near-IR spectrograph Keck II/NIRSPEC \citep{McLean1998}  at a higher resolution  (FWHM $\approx 240~\rm km\, s^{-1}$) than achieved by LRIS-B \citep[see][for details on this galaxy sample]{Erb2006}. The sample used here is part of a larger sample, with a few thousand galaxies, from which we selected all galaxies within 2 Mpc from the LOS to the background QSOs, and with redshifts that fall in the Ly$\alpha$ forest redshift ranges of the QSOs in their fields. We define the Ly$\alpha$ forest range as the part of the spectrum between the quasar's Ly$\alpha$ and Ly$\beta$ emission lines, excluding the region $<5000\, \rm km\, s^{-1}$ from the quasar's redshift (to avoid the proximity zone).

The QSO observations were conducted between 1996 and 2009 with the Keck I/HIRES echelle spectrograph. The spectra have a resolution of $\Delta v\approx7.5\, \rm km\, s^{-1}$ ($\rm R\simeq 40,000$) and were rebinned into pixels of 0.04\,\AA. They cover  the wavelength range  $\approx3100 - 6000$\,\AA. The reduction was done using T. Barlow's MAKEE package, and the spectra were normalized using low order spline fits. The typical S/N in the Ly$\alpha$ forest region lies between 40 and 160.

Repeated observations of the same galaxies suggest that the typical statistical measurement uncertainties are $\approx 100~\rm km\, s^{-1}$ for redshifts measured from ISM absorption lines ($z_{\rm ISM}$),  $\approx 50~\rm km\, s^{-1}$ for  redshifts measured from Ly$\alpha$ emission lines ($z_{\rm Ly\alpha}$), and  $ \approx 60~\rm km\, s^{-1}$ for  redshifts measured from nebular emission lines ($z_{\rm neb}$). The systematic offsets between these redshifts and the true (i.e.\ systemic) redshifts, which is the subject of this paper, may of course be larger.

\section{Method}\label{Method}
\begin{figure}
\resizebox{\colwidth}{!}{\includegraphics{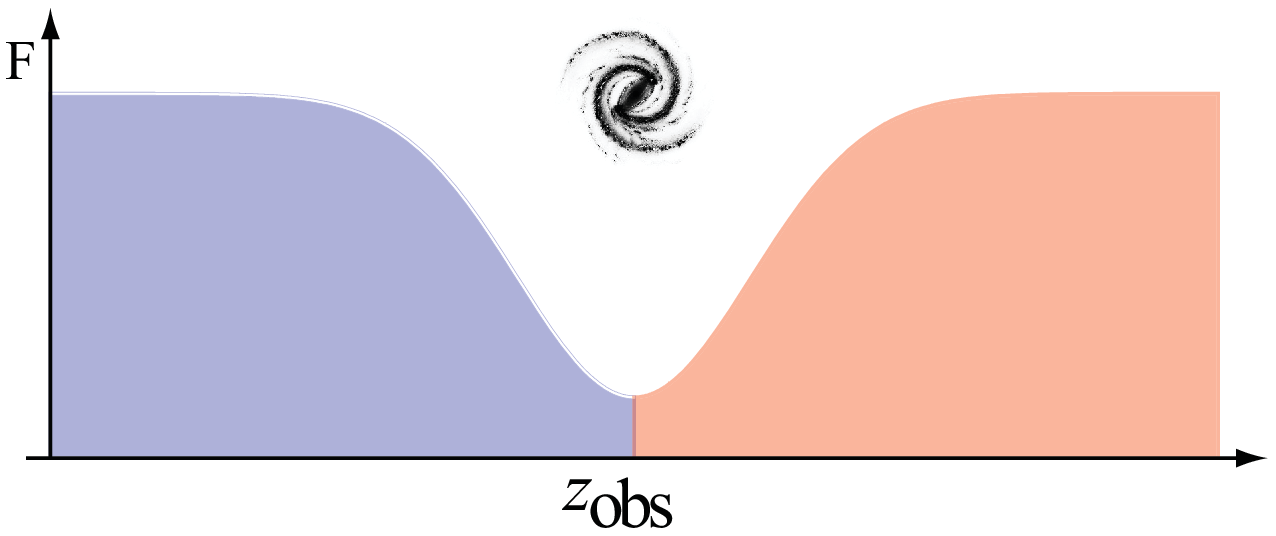}\includegraphics{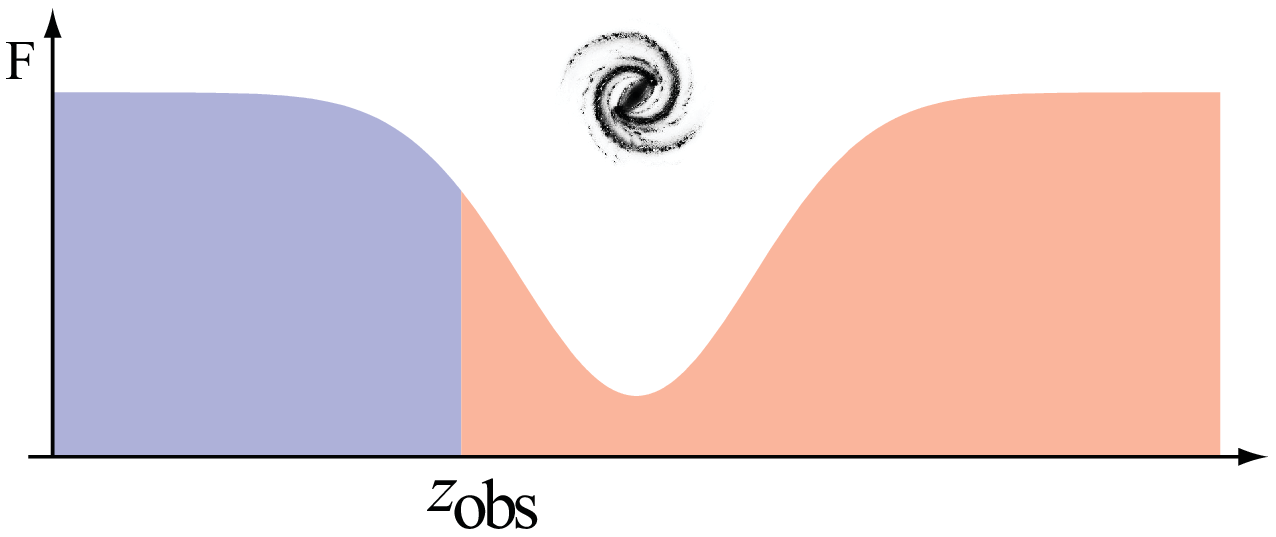}}
\caption{The mean flux profile is symmetric around galaxy positions if the observed redshifts ($z_{\rm obs}$, vertical red line) are equal to the true redshifts (left panel). If the observed redshifts are systematically offset from the true redshifts, the mean profile will be symmetric around some other redshift point, offset from the observed redshifts (right panel).}\label{drawing}
\end{figure}

\indent Absorption spectra of background objects (QSOs, GRBs, or star-forming galaxies) can be used to study the intervening IGM. Observations generally show an increase in the mean Ly$\alpha$ and metal absorption as the distance to the nearest galaxy decreases. One expects to find an average absorption profile that is symmetric around the galaxies'  positions along the LOS. This is true even if the IGM is not distributed  isotropically around galaxies, which can for example  occur if the mean absorption at a given distance from a galaxy correlates with the galaxy's inclination. Assuming that the galaxies are oriented randomly  with respect to the LOS, we still expect to see a symmetric absorption profile when averaged over an ensemble of galaxies\footnote{As the blue side corresponds to a lower Ly$\alpha$ redshift than the red side, and the mean Ly$\alpha$ absorption at a random location is known to increase with redshift, we do expect the intergalactic Ly$\alpha$ absorption to be slightly weaker on the blue side. However, the magnitude of this effect \citep[the mean transmission varies by 0.002 over $10^3\rm \, km\, s^{-1}$ at $z=2.3$; e.g.\ ][]{Schaye2003} is negligible compared to the enhancement in the mean absorption near galaxies.}.  If the measured redshifts of galaxies are systematically offset from their systemic redshifts, then the mean profile will be symmetric around some other velocity point, offset from the observed redshift by an amount equal to the systematic shift in galaxy redshifts  (see Fig.~\ref{drawing} for an illustration). In this paper we use this principle to calibrate the systematic offsets between the observed and systemic (i.e. true) redshifts for the galaxies in our sample.

One could imagine that instead of measuring the offset from the average absorption profile, we could measure offsets for individual galaxies by observing the IGM absorption features around them, and then find the average of such estimated offsets. However, near the redshift of  a galaxy one often finds several lines at different velocities, of which some are associated with the galaxy in question, but the others are produced by, for example, a cloud that is associated with a neighboring galaxy or an intervening cloud. If the absorption line associated with such a cloud were stronger than that due to the gas associated with the galaxy, then we would infer the wrong offset. Also, at small impact parameters even a cloud that is associated with the galaxy in question could potentially  have kinematics that is not dominated by gravity (e.g., it could be a part of the galactic outflow); and at any point in the forest there could be absorption completely unrelated to the observed redshift. This is not a problem when using the average absorption profile, given that it is equally likely for such a line to be at a positive  as at a negative velocity with respect to the galaxy.

We use Ly$\alpha$ rather than metal lines, because the metal-line absorption is usually only enhanced within a couple of hundred kpc from galaxies, and our sample does not include a sufficiently large number of galaxies with such small impact parameters  (see the discussion below on the required number of galaxies). However, at higher redshifts, where the Ly$\alpha$ forest starts to saturate, it might be advantageous  to use metal  lines, as long as the survey contains sufficient galaxies.
 
We compare the ``blue'' and ``red'' sides of the median Ly$\alpha$ absorption profile, within $\pm 500\, \rm km\, s^{-1} $ of the galaxy redshifts of all the foreground galaxies with impact parameters, $b$, smaller than 2 Mpc. We try a large number of velocity offsets, $z_{\rm obs} \rightarrow z_{\rm obs}+\frac{\Delta v}{c}(1+z_{\rm obs})$, where  $z_{\rm obs}$ is the observed redshift, and $c$ is the speed of light. Note that a positive $\Delta v$ indicates that the observed redshifts systematically underestimate the true redshifts.We verified that there is no justification for using a redshift dependent offset -- either because our redshift range is too small for evolution to matter ($z\approx1.9-2.6$), or because there is no dependence on redshift in this galaxy population.

The analysis involves the following steps: \it{i)} \normalfont we assume that the observed redshifts are systematically offset from the systemic redshifts by $\Delta v$ and assign each galaxy a new redshift $z'_{\rm obs}=z_{\rm obs}+\frac{\Delta v}{c}(1+z_{\rm obs})$; \it{ii)} \normalfont for each galaxy we shift the QSO spectrum into the galaxy's rest-frame; \it{iii)} \normalfont for each galaxy we interpolate the flux in the QSO spectrum within $\pm500\, \rm km\, s^{-1}$ from $z'_{\rm obs}$ in velocity bins of $10\, \rm km\, s^{-1}$ to get an absorption profile; \it{iv)} \normalfont we find the median\footnote{We verified that using the mean rather than the median gives nearly identical results for samples with enough galaxies, but that using the median is more robust for small samples.} flux profile as a function of velocity separation from galaxies; \it{v)} \normalfont  we compare the ``blue'' ($-500 - 0\, \rm km\, s^{-1}$) and the ``red'' ($0-500\, \rm km\, s^{-1}$) side of the median flux profile using $\chi^{2}$ statistics: 
$$
\chi^{2}=\sum_{i=1}^{N}\frac{\rm (B_{i}-R_{i})^{2}}{\rm \sigma^{2}(B_{i})+\sigma^{2}(R_{i})}
$$
where B$_{i}$ and R$_{i}$ are the blue and red median flux arrays, $\sigma(\rm B_{i})$ and $\sigma(\rm R_{i})$ are the errors on the median flux, and the sum extends over all velocity bins within $\pm 500\, \rm km\, s^{-1}$ from the assumed redshifts.  The errors on the median flux profiles are estimated by bootstrap resampling. We recreated the galaxy sample 1000 times by dividing each QSO field into redshift bins of $1000\rm \, km\, s^{-1}$ and then randomly drawing redshift intervals to recreate each field, allowing for individual intervals to be selected multiple times. Each bootstrapped galaxy sample consists of galaxies whose redshifts fall in the selected redshift intervals. We chose an interval of $1000\rm \, km\, s^{-1}$ because this length is larger than the correlation length of LBGs \citep{Adelberger2005clustering}. These five steps are repeated for a large number of assumed redshift offsets, yielding a reduced $\chi^2$ curve (see Fig.~\ref{EMISM1D}). We expect the $\chi^{2}$  to reach a (local) minimum for the offset that brings the measured redshifts closest to the true redshifts.  Note that the $\chi^{2}$ will also have a small value if the applied redshift offsets are very large, because in that case the absorption will no longer be correlated with the assumed locations of galaxies, the median flux profile will be flat, and the blue and red sides will agree. We therefore search for  $\chi^{2}$ minima only between the inferred $\chi^{2}$ maxima (see Fig.~\ref{EMISM1D} for the  characteristic shape of the $\chi^{2}$ curves). In practice we do this as follows: \emph{i)} we smooth the $\chi^{2}$ curve with a boxcar average with $\rm 50\, km\, s^{-1}$ width; \emph{ii)} we find the global $\chi^{2}$ maximum and determine the FWHM of the $\chi^2$ peak it belongs to; \emph{iii)} we search for the second (local) $\chi^{2}$ maximum that is at least one FWHM away from the first one; \emph{iv)} we find the $\chi^{2}$ minimum between these two peaks, in the un-smoothed curve.

The value that we chose for the maximum impact parameter, $\rm b=2\, Mpc$, is a compromise between the need for a strong absorption signal (the mean absorption is stronger for galaxies with smaller impact parameters), and the need for a large number of galaxies (there are more galaxies with larger impact parameters). We verified that using smaller maximum  impact parameters gives consistent results. Any maximum impact parameter in the range\footnote{There are insufficient galaxies with $b<0.5\rm \, Mpc$ to test smaller maximum impact parameters.} 0.5-2 Mpc yields redshift calibrations that agree with those obtained for the $b<2\, \rm Mpc$ sample within the estimated errors. However, the required number of galaxies does depend on the adopted maximum impact parameter. For $b<1$ and $b<2\, \rm Mpc$, we obtain converged results if we use 
$\ga100$ and $\ga200$ galaxies, respectively, although the error bar on the inferred systematic shift continues to decrease slowly if more galaxies are included. We also varied the velocity range around each galaxy. In this case there is a tradeoff between the strength of the signal and the number of pixels available for statistics. We found that the velocity range of $\pm500\, \rm km\, s^{-1}$ generally gives the tightest confidence intervals around the estimated offsets, but the results are insensitive to the precise velocity range chosen.

We  apply the above method to our sample and estimate systematic velocity shifts for the subsamples listed in Fig.~\ref{ComparisonExclMS_Short}. We make a distinction between the whole sample and the NIRSPEC subsample due to the possibility that this subset is not representative of the whole sample \citep[see][for details on the sample selection and possible biases]{Erb2006}.

\section{Results}
\begin{figure*}
\resizebox{\colwidth}{!}{\includegraphics{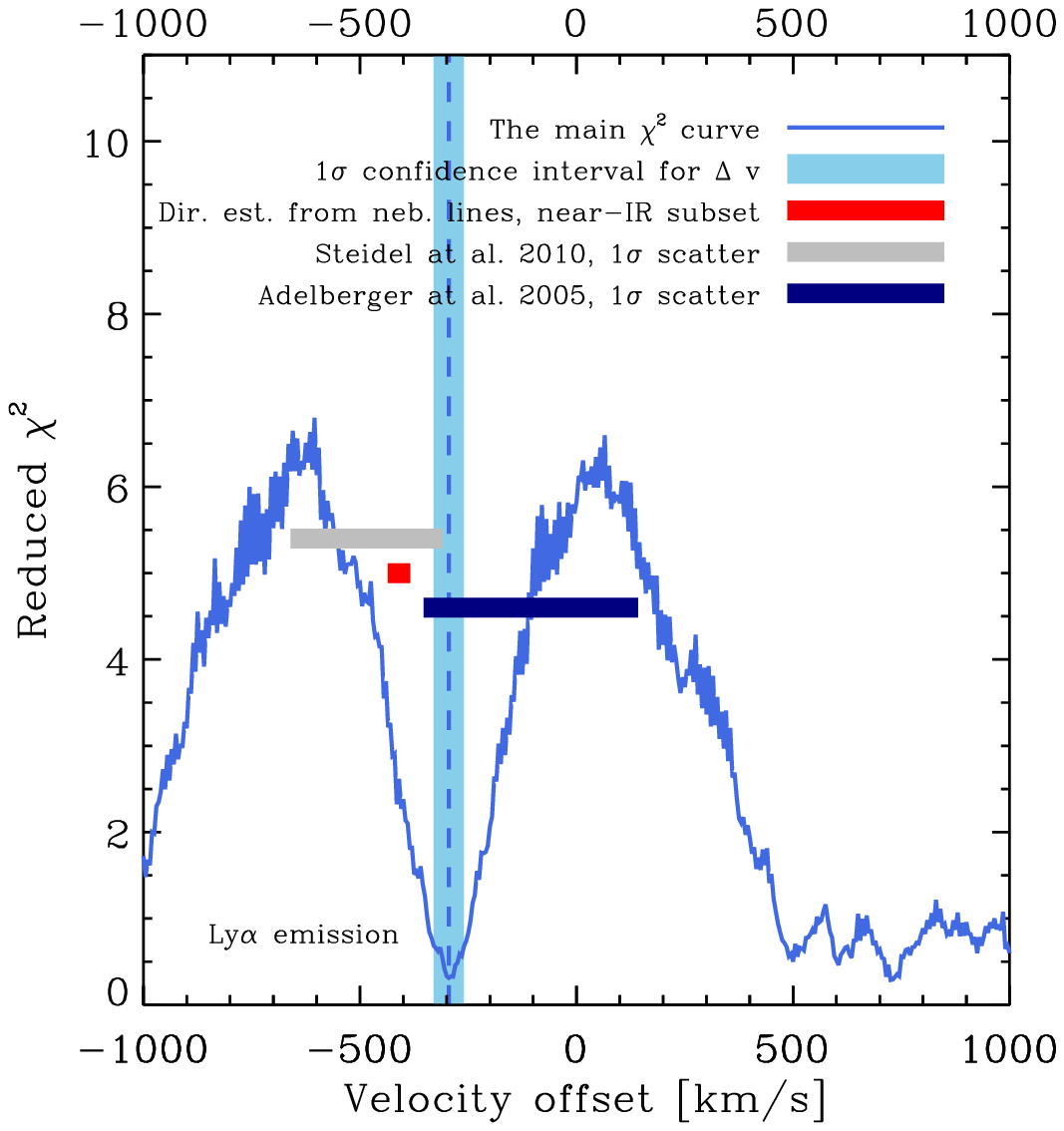}}
\resizebox{\colwidth}{!}{\includegraphics{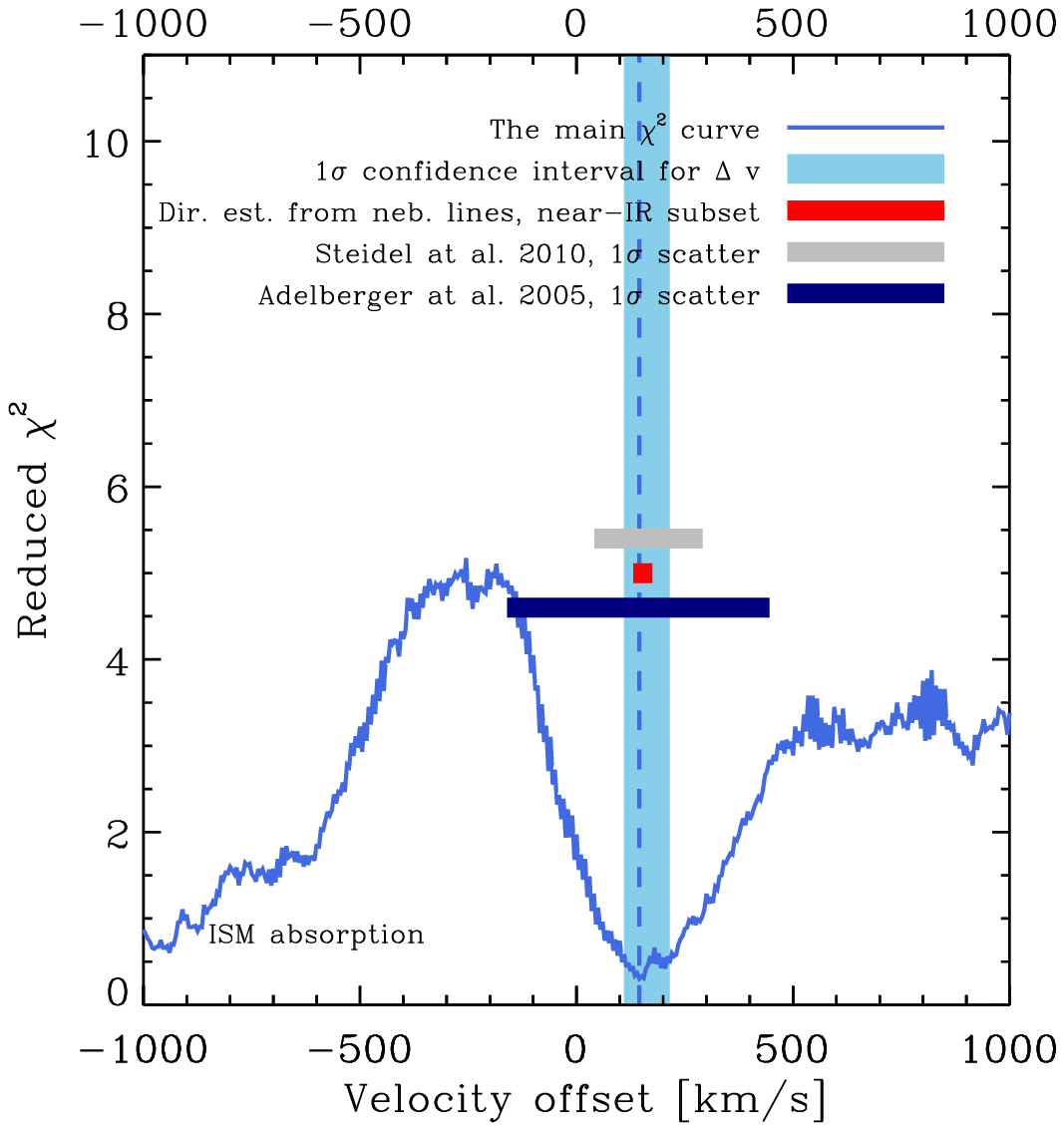}}
\caption{The reduced $\chi^{2}$ obtained by comparing the red and blue sides of the median Ly$\alpha$ absorption profile, as a function of assumed velocity offset $\Delta v=c(z_{\rm gal}-z_{\rm obs})/(1+z_{\rm obs})$, where $z_{\rm gal}$ and $z_{\rm obs}$ are the true systemic and observed redshifts, respectively. The red and the blue sides are compared within $\pm500 \rm\,km\ s^{-1}$ of all galaxies with impact parameters smaller than 2 Mpc. The \emph{left panel}  shows the results for redshifts measured from \emph{Ly$\alpha$ emission lines}  (321 galaxies), and the \emph{right panel}  for redshifts measured from \emph{ISM absorption lines}   (590 galaxies). The  dashed vertical line shows the (local) minimum of the $\chi^{2}$ curve.  The vertical, light blue regions show the 1$\sigma$ confidence intervals, as estimated by bootstrap resampling the galaxy samples. The $\chi^{2}$ becomes small  for very large offsets because in that case we are  comparing ``no signal with no signal'' since we completely miss the galaxies. The horizontal red polygons show the 1$\sigma$ confidence interval for the median offset for the near-IR subsample, obtained by direct comparison with nebular lines \citep[see also][]{Steidel2010}. The grey and dark blue horizontal lines show the estimates for the mean offset from \citet{Steidel2010} and \citet{Adelberger2005} obtained from nebular lines, respectively, with the size of the bars indicating the 1$\sigma$ scatter among the galaxies. Our measurements agree well with direct estimates based on comparison with nebular lines for ISM absorption lines, but for Ly$\alpha$ emission lines they differ significantly.}\label{EMISM1D}
\end{figure*}

\indent In Fig.~\ref{EMISM1D} we plot the reduced $\chi^{2}$ as a function of the assumed velocity offset. We find that the best offset for  redshifts measured from Ly$\alpha$ emission is $\Delta v_{\rm Ly\alpha}=-295_{-35}^{+35}\rm \, km\, s^{-1}$ (i.e., the Ly$\alpha$ emission line is systematically redshifted with respect to the systemic velocity),  while for redshifts measured from ISM absorption lines we find $\Delta v_{\rm ISM}=145_{-35}^{+70}\rm\, km\, s^{-1}$ (i.e., the ISM absorption lines are systematically blueshifted).

We estimated the confidence intervals around the best estimates for $\Delta v_{\rm ISM}$ or $\Delta v_{\rm Ly\alpha}$ using the bootstrap method, described in Section~\ref{Method}. We perform all the steps described in Section~\ref{Method} on  samples created by bootstrapping the original sample, resulting in a best-fit $\Delta v$ for each bootstrap realization of the galaxy sample. The error on the estimate of $\Delta v$ is the $1\sigma$ confidence  interval around the median $\Delta v$ from the 1000 bootstrap realizations.  These 1$\sigma$ confidence intervals  are shown as the light blue shaded regions in Fig.~\ref{EMISM1D} together with the $\chi^{2}$ curve.

We prefer to determine confidence intervals by using bootstrap resampling rather than $\Delta\chi^{2}$ because the former is more robust. Indeed, we found that for small galaxy samples the $\chi^{2}$ curves become too noisy for $\Delta\chi^{2}$ estimates to work, while for large samples they yield errors slightly larger than those estimated from bootstrap resampling. Contrary to $\Delta\chi^{2}$, bootstrap resampling does not require the errors on the median flux profile to be Gaussian. It also does not require the different velocity bins to provide independent measurements, which they in fact do not because the individual absorption lines are broader than the bins in $\Delta v$. Bootstrap resampling merely requires that the redshift regions into which we divide the fields are independent, which is true since, as already mentioned, they are larger than the galaxies' correlation length. We thus only use the $\chi^{2}$ curve to find the $\Delta v$ that minimizes the difference between the blue and red sides.

For the subset of galaxies with near-IR data, the nebular lines provide accurate markers of the systemic redshifts of the individual galaxies. Comparing these with the rest-frame UV lines, we find median offsets of  $\Delta v_{\rm Ly\alpha}=-406_{-30}^{+22}\rm \, km\, s^{-1}$ and $\Delta v_{\rm ISM}=166_{-35}^{+9}\rm\, km\, s^{-1}$  (based on 42 and 86 galaxies, respectively, including objects that are not within 2~Mpc of the LOS to a QSO). The quoted uncertainties correspond to the 1$\sigma$ confidence interval around  the median, estimated by bootstrap resampling the galaxy sample 1000 times. These results are shown as red polygons in Fig.~\ref{EMISM1D}. For reference, we note that \citet{Steidel2010} quote\footnote{\citet{Steidel2010} also quote $\Delta v_{\rm Ly\alpha}=-485\pm 175\rm \, km\, s^{-1}$ and $\Delta v_{\rm ISM}=166\pm{125}\rm\, km\, s^{-1}$, but these errors  are standard deviations around the best offset estimates.} mean offsets of  $\Delta v_{\rm Ly\alpha}=-445\pm 27\rm \, km\, s^{-1}$ and $\Delta v_{\rm ISM}=164\pm{16}\rm\, km\, s^{-1}$. While our result for $\Delta v_{\rm ISM} $ agrees with the direct estimate,  the best estimates for $\Delta v_{\rm Ly\alpha}$ differ significantly.

We therefore employ our IGM calibration method for the subsample with nebular redshifts. For Ly$\alpha$ the resulting offset is $\Delta v_{\rm Ly\alpha}=-455_{-65}^{+45}\rm \, km\, s^{-1}$. This is in good agreement with the direct estimate  ($\Delta v_{\rm Ly\alpha}=-406_{-30}^{+22}\rm \, km\, s^{-1}$) from nebular lines for the same galaxies. Both disagree with our measurement for the full sample ($\Delta v_{\rm Ly\alpha}=-295_{-35}^{+35}\rm \, km\, s^{-1}$). 

We emphasize that our measurement of $\Delta v_{\rm Ly\alpha}$ for the near-IR sample is based on only 26 galaxies, which is generally insufficient to obtain a converged result. Indeed, we find that our method fails for the subsample of 48 galaxies with both ISM absorption and nebular emission lines. The reduced $\chi^2$ curve is noisy and does not show two clear peaks, which causes the best-fit velocity offset to vary strongly with the velocity interval over which we measure the median flux. Although the $\chi^2$ curve for the subsample of near-IR galaxies with Ly$\alpha$ emission does look normal and we found the result to be insensitive to the precise velocity interval used, the number of galaxies may well be too small to obtain a robust measurement. 

Galaxies whose redshifts have been measured only from their Ly$\alpha$ emission feature tend to be those with poorer continuum S/N, due either to faintness or to poor observing conditions. Such galaxies are under-represented in the near-IR subsample because their redshifts are considered less secure and therefore
were selected against for near-IR spectroscopic follow-up \citep{Erb2006}. It has also been observed that the velocity offset of Ly$\alpha$ emission is anti-correlated with the Ly$\alpha$ equivalent width \citep{Shapley2003} -- stronger Ly$\alpha$ lines are expected to have velocities closer to the galaxy systemic redshift. These two effects are likely responsible for the differences between the full sample and the smaller near-IR subsample.

This difference sends a warning that describing a galaxy population with a single number is not the best strategy given that the line offsets depend on galaxy spectral morphology. Thus, conclusions drawn from the near-IR subsample may not be generally applicable and separate redshift calibrations are needed  for each sample of interest.

However, we do confirm the finding from \citet{Steidel2010} that absorption line offsets are less sensitive to galaxy spectral morphology than Ly$\alpha$ emission line offsets. This justifies the strategy taken in \citet{Adelberger2003,Adelberger2005} and \citet{Steidel2010} to use redshifts measured from ISM absorption lines when they are available ($\approx90\%$ of the sample).

\begin{figure*}
{\includegraphics[height=3in]{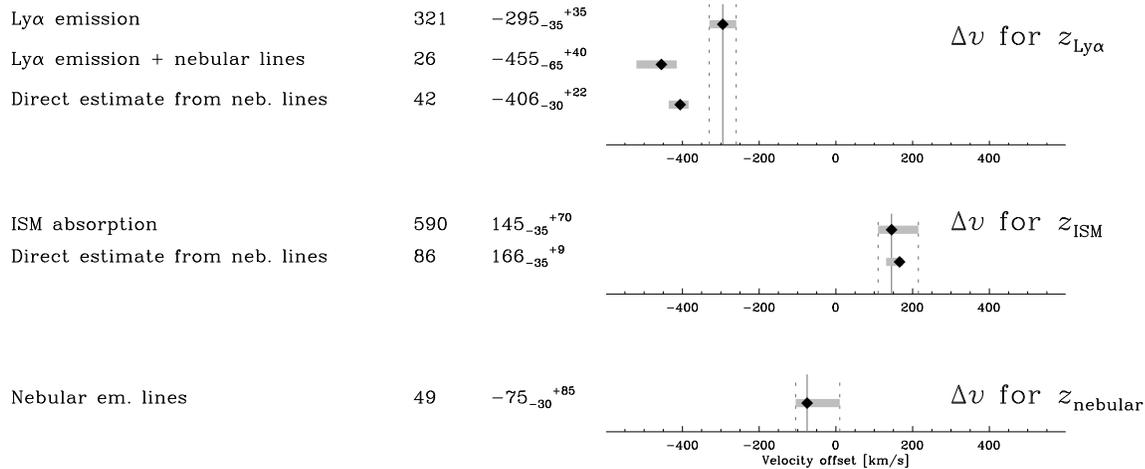}}
\caption{Different subsamples (first column) together with the number of galaxies in the sample (second column) and the best-fit velocity offsets, $\Delta v$, and  $1\sigma$ confidence intervals (third column). These best offsets (black diamonds) together with the $1\sigma$ confidence intervals (grey horizontal stripes) are shown graphically on the right. The results for the full samples are  shown as vertical gray lines. }\label{ComparisonExclMS_Short}
\end{figure*}

As an additional test of our IGM calibration method, we measure the systematic velocity offset for redshifts measured from nebular emission lines, which are known to be close to systemic positions. The resulting offset, which is based on a sample of 49 galaxies, is  $\Delta v_{\rm nebular}=-75_{-30}^{+85}\rm\, km\, s^{-1}$, i.e. consistent with zero, as expected (Fig.~\ref{ComparisonExclMS_Short}).

\subsection{Random offsets}
\indent We have demonstrated  that one can use the fact that Ly$\alpha$ absorption by the IGM is correlated with proximity to galaxies  to calibrate the redshifts inferred from Ly$\alpha$ emission and ISM absorption lines. In other words, we measured the systematic offsets in the observed redshifts. We will now use the same idea to constrain  the statistical (i.e.\ random) offsets. Statistical scatter in individual galaxy redshift offsets may be caused by measurement and instrumental errors, but in our case it is probably dominated by the variations in the  intrinsic galaxy properties. By comparing redshifts inferred from rest-frame UV lines with those from nebular lines for the subset of 89 galaxies that have been observed in the near-IR, one not only finds a large systematic offset, but there is also a significant scatter of $\approx 180\, \rm km\, s^{-1}$ for Ly$\alpha$ redshifts, and $\approx 160\, \rm km\, s^{-1}$ for ISM redshifts. 

We use all galaxies with impact parameters smaller than 2 Mpc, and measure the median absorption around them. We then repeat the procedure after applying a random redshift offset to
the redshift of each galaxy, drawn from a Gaussian distribution with mean zero and standard deviation $\sigma$. We do this many times for  increasing values of $\sigma$. We then compare the resulting median flux profiles with  the original (i.e.\ $\sigma=0$) median absorption profile. We expect that the two profiles will be consistent with each other as long as the added random offsets are small compared with the true random offsets. Using this method we obtain 1$\sigma$ upper limits on the random redshift offsets of $<220\, \rm km\, s^{-1}$ ($<300\, \rm km\, s^{-1}$) for Ly$\alpha$ redshifts, using mean (median) statistics, and  $<420\, \rm km\, s^{-1}$ ($430\, \rm km\, s^{-1}$) for ISM redshifts, which is consistent with the independent estimates based on the subsample with redshifts from nebular lines. The quoted values are upper limits because we measure the amount of scatter needed to cause a significant difference with respect to the original absorption profile, i.e. the value we find exceeds the actual scatter that is present in the sample.

\section{Summary \& Conclusions}
Measuring accurate redshifts for high-$z$ galaxies is a daunting task. Distant objects are faint, making even ground-based rest-frame UV observations challenging. Added to this is the complexity of the gas kinematics and radiation transport, which may cause absorption and emission features to be systematically offset from the true systemic redshifts. Here we have used a sample of 15 absorption spectra of bright background QSOs, with more than 600 foreground galaxies within 2 Mpc from the LOS to these QSOs, to probe the IGM close to galaxies at $z\approx2-2.5$. All galaxies have spectroscopic redshifts measured from either the \ion{H}{I} Ly$\alpha$ emission line and/or rest-frame UV ISM absorption lines. In addition, for a subset of 49 galaxies redshifts have been measured from rest-frame optical nebular emission lines, which are thought to be unbiased with respect to the systemic redshifts. By utilizing the fact that the mean absorption profiles must be symmetric with respect to the true galaxy redshifts if the galaxies are oriented randomly with respect to the LOS,   we calibrated the redshifts measured from rest-frame UV spectral lines. 

Our results are summarized in Fig.~\ref{ComparisonExclMS_Short}. We found that observed  Ly$\alpha$ emission redshifts require a systematic shift of $\Delta v_{\rm Ly\alpha}=-295_{-35}^{+35}\rm \, km\, s^{-1}$ (i.e.\ the Ly$\alpha$ lines are redshifted with respect to the systemic velocity) and that ISM absorption redshifts need an offset of  $\Delta v_{\rm ISM}=145_{-35}^{+70}\rm\, km\, s^{-1}$ (i.e.\ the ISM lines are blueshifted). For the nebular lines we found $\Delta v_{\rm
nebular}=-75_{-30}^{+85}\rm\, km\, s^{-1}$, which is consistent with zero, as expected.

We compared our calibrations to those obtained by direct comparison with nebular emission lines, which should be close to systemic, for a subsample observed with NIRSPEC/Keck. While our measurements for $\Delta v_{\rm ISM}$ agree, the direct measurement for the systematic offset of the Ly$\alpha$ emission line, $\Delta v_{\rm Ly\alpha}=-406_{-30}^{+22}\rm \, km\, s^{-1}$, differs significantly. However, when we applied our method to the NIRSPEC subsample we found $\Delta v_{\rm Ly\alpha}=-455_{-65}^{+45}\rm \, km\, s^{-1}$, in agreement with the direct measurement. The discrepancy between the Ly$\alpha$ offsets between the near-IR subsample and the full sample reflects the fact that the line offset depends on galaxy spectral morphology, which is different for the NIRSPEC subsample due to the way in which these galaxies were selected. 

After having demonstrated  that one can use IGM absorption to measure systematic galaxy redshift offsets, we also estimated upper limits on the random offsets for individual galaxies. In our case these random offsets are likely dominated by intrinsic scatter rather than by measurement errors. We applied random velocity shifts  drawn from a Gaussian distribution of varying width to the 
redshift of each galaxy until the median Ly$\alpha$ absorption profile close to galaxies was significantly affected. This procedure yielded 1$\sigma$ upper limits on the random offsets of $<220\, \rm km\, s^{-1}$ ($<300\, \rm km\, s^{-1}$) for Ly$\alpha$ redshifts, using mean (median) statistics, and  $<420\, \rm km\, s^{-1}$ ($<430\, \rm km\, s^{-1}$) for ISM redshifts, which are consistent with direct measurements for the subsample with nebular redshifts. 

These results will for example be of interest for future Ly$\alpha$ emitter surveys for which  follow-up near-IR spectroscopy will not be available. Such surveys would otherwise have to rely on previous calibrations for LBGs at $z\approx 2-3$, which might  not be appropriate for different galaxy populations or redshifts. Without applying our self-calibration technique, the redshifts based on Ly$\alpha$ would likely be systematically wrong by a few hundred $\rm km\, s^{-1}$, which would  make it very difficult to study the galaxies' environments through absorption line spectroscopy of nearby background QSOs or GRBs. 

In future papers we will also use the calibrations presented here to study the IGM near the galaxies in our sample.

\section*{Acknowledgments}
We are very grateful to Alice Shapley, Dawn Erb, Naveen Reddy, Milan Bogosavljevi\' c, and Max Pettini for their help with collecting and processing the data. We also thank the anonymous referee whose comments improved the clarity of the paper. This work was supported by an NWO VIDI grant (O.R., J.S.), by the US National Science Foundation through grants AST-0606912 and AST-0908805, and by the David and Lucile Packard Foundation (C.C.S.). C.C.S. acknowledges additional support from the John D. and Catherine T. MacArthur Foundation and the Peter and Patricia Gruber Foundation. We thank the W. M. Keck Observatory staff  for their assistance with the observations. We also thank the Hawaiian people, as without their hospitality the observations presented here would not have been possible.

\label{lastpage}


\begin{thebibliography}{99}
\bibitem[\protect\citeauthoryear{Adelberger et al.}{2003}]{Adelberger2003} Adelberger K. L.,  Steidel C. C., Shapley A. E., Pettini M., 2003, \apj, 584, 45
\bibitem[\protect\citeauthoryear{Adelberger et al.}{2005a}]{Adelberger2005clustering} Adelberger K.~L., Steidel C.~C., Pettini 
M., Shapley A.~E., Reddy N.~A., Erb D.~K., 2005a, \apj, 619, 697 
\bibitem[\protect\citeauthoryear{Adelberger et al.}{2005b}]{Adelberger2005} Adelberger K. L., Shapley A. E., Steidel C. C., Pettini M., Erb D. K., Reddy N. A., 2005b, \apj, 629, 636
\bibitem[\protect\citeauthoryear{Bacon et al.}{2010}]{Bacon2010} {Bacon} R. et al., 2010, Proc. SPIE,  7735, 7
\bibitem[\protect\citeauthoryear{Bergeron \& Boiss\'e}{1991}]{Bergeron1991} Bergeron J., Boiss\'e P., 1991, \aap, 243, 344 
\bibitem[\protect\citeauthoryear{Bowen et al.}{2002}]{Bowen2002}{Bowen} D.~V., {Pettini} M.,  {Blades} J.~C., 2002, \apj, 580, 169
\bibitem[\protect\citeauthoryear{Chen et al.}{1998}]{Chen1998}{Chen} {H.-W.}, {Lanzetta} K.~M., {Webb} J.~K., {Barcons} X., 1998, \apj, 498, 77C
\bibitem[\protect\citeauthoryear{Chen et al.}{2001}]{Chen2001}{Chen} {H.-W.}, {Lanzetta} K.~M., {Webb} J.~K., 2001, \apj, 556, 158
\bibitem[\protect\citeauthoryear{Crighton et al.}{2010}]{Crighton2010} Crighton N. et al., 2010 \mnras, in press (astro-ph/1006.4385)
\bibitem[\protect\citeauthoryear{Dijkstra et al.}{2006a}]{Dijkstra2006a}{Dijkstra} M., {Haiman} Z., {Spaans} M., 2006a, \apj, 649, 14
\bibitem[\protect\citeauthoryear{Dijkstra et al.}{2006b}]{Dijkstra2006b}{Dijkstra} M., {Haiman} Z., {Spaans} M., 2006b, \apj, 649, 37
\bibitem[\protect\citeauthoryear{Erb et al.}{2006}]{Erb2006} Erb D. K., Steidel C. C., Shapley A. E.,  Pettini M., Reddy N. A., Adelberger K. L., 2006, \apj, 646, 107
\bibitem[\protect\citeauthoryear{Hansen \& Oh}{2006}]{Hansen2006}{Hansen} M.,  {Oh} S.~P., 2006, \mnras, 367, 979
\bibitem[\protect\citeauthoryear{Komatsu et al.}{2009}]{Komatsu2009} {Komatsu} E. et al., 2009, \apjs, 180, 330
\bibitem[\protect\citeauthoryear{Lanzetta \& Bowen}{1990}]{Lanzetta1990}{Lanzetta} K.~M., {Bowen} D., 1998, \apj, 357, 321
\bibitem[\protect\citeauthoryear{Lanzetta et al.}{1995}]{Lanzetta1995} {Lanzetta} K.~M., {Bowen} D.~V., {Tytler} D., {Webb} J.~K., 1995, \apj, 442, 538
\bibitem[\protect\citeauthoryear{McLean et al.}{1998}]{McLean1998} {McLean} I.~S. et al., 1998,  Proc. SPIE, 3354, 566
\bibitem[\protect\citeauthoryear{McLean et al.}{2008}]{McLean2008} {McLean} I.~S., {Steidel} C.~C., {Matthews} K., {Epps} H., {Adkins} S.~M., 2008, Proc. SPIE, 7014, 99
\bibitem[\protect\citeauthoryear{Morris \& Jannuzi}{2006}]{Morris2006} Morris S.~L., Jannuzi B.~T., 2006, \mnras, 367,1261
\bibitem[\protect\citeauthoryear{Penton et al.}{2002}]{Penton2002}{Penton} S.~V., {Stocke} J.~T., {Shull} J.~M., 2002, \apj, 565, 720
\bibitem[\protect\citeauthoryear{Pettini et al.}{2001}]{Pettini2001} Pettini M., Shapley A.~E., Steidel C.~C., Cuby J.-G., Dickinson M., Moorwood A.~F.~M., Adelberger K.~L., Giavalisco M., 2001, \apj, 554, 981
\bibitem[\protect\citeauthoryear{Pieri et al.}{2006}]{Pieri2006}{Pieri} M.~M., {Schaye} J., {Aguirre} A., 2006, \apj, 638, 45
\bibitem[\protect\citeauthoryear{Reddy et al.}{2005}]{Reddy2005} {Reddy} N.~A., {Erb} D.~K., {Steidel} C.~C., {Shapley} A.~E., {Adelberger} K.~L., {Pettini} M., 2005, \apj, 633, 748
\bibitem[\protect\citeauthoryear{Rupke et al.}{2002}]{Rupke2002} Rupke D.~S., Veilleux S., Sanders D.~B., 2002, \apj, 570, 588
\bibitem[\protect\citeauthoryear{Rupke et al.}{2005}]{Rupke2005} Rupke D.~S., Veilleux S., Sanders D.~B., 2005, \apjs,  160, 115
\bibitem[\protect\citeauthoryear{Ryan-Weber}{2006}]{Ryan-Weber2006}  Ryan-Weber E.~V., 2006, \mnras, 367, 1251
\bibitem[\protect\citeauthoryear{Schaerer \& Verhamme}{2008}]{Shaerer2008}{Schaerer} D., {Verhamme} A., 2008, \aap, 480, 369
\bibitem[\protect\citeauthoryear{Schaye et al.}{2003}]{Schaye2003} {Schaye} J., {Aguirre} A., {Kim} {T.-S.}, {Theuns} T., 
	{Rauch} M.,  {Sargent} W.~L.~W., 2003, \apj, 596, 768
\bibitem[\protect\citeauthoryear{Shapley et al.}{2003}]{Shapley2003} Shapley A.~E., Steidel C.~C., Pettini M., Adelberger K.~L., 2003, \apj, 588, 65
\bibitem[\protect\citeauthoryear{Steidel et al.}{1994}]{Steidel1994} {Steidel} C.~C., {Dickinson} M., {Persson} S.~E., 1994, \apj, 437, 75
\bibitem[\protect\citeauthoryear{Steidel et al.}{1997}]{Steidel1997} {Steidel} C.~C., {Dickinson} M., {Meyer} D.~M., {Adelberger} K.~L., {Sembach} K.~R., 1997, \apj, 480, 568
\bibitem[\protect\citeauthoryear{Steidel et al.}{2004}]{Steidel2004} {Steidel} C.~C., {Shapley} A.~E., {Pettini} M., {Adelberger} K.~L., {Erb} D.~K., {Reddy} N.~A., {Hunt} M.~P., 2004, \apj, 604, 534
\bibitem[\protect\citeauthoryear{Steidel et al.}{2010}]{Steidel2010} {Steidel} C.~C., {Erb} D.~K., {Shapley} A.~E., {Pettini} M., {Reddy} N., {Bogosavljevi{\'c}} M., {Rudie} G.~C., {Rakic} O., 2010, \apj, 717, 289 
\bibitem[\protect\citeauthoryear{Verhamme et al.}{2006}]{Verhamme2006}{Verhamme} A., {Schaerer} D., {Maselli} A., 2006, \aap, 460, 397
\bibitem[\protect\citeauthoryear{Verhamme et al.}{2008}]{Verhamme2008}{Verhamme} A., {Schaerer} D., {Atek} H., {Tapken} C., 2008, \aap, 491, 89
\bibitem[\protect\citeauthoryear{Wilman et al.}{2007}]{Wilman2007} Wilman R.~J., Morris S.~L., Jannuzi B.~T.,  Dav{\'e} R., Shone A.~M., 2007, \mnras, 375, 735
\bibitem[\protect\citeauthoryear{Zheng \& Miralda-Escud{\'e}}{2002}]{Zheng2002}{Zheng} Z., {Miralda-Escud{\'e}} J., 2002, \apj, 578, 33
\end{thebibliography}
\end{document}